\documentclass[preprint,showpacs,preprintnumbers,amsmath,amssymb]{revtex4}
\usepackage{graphicx}
\usepackage{dcolumn}
\usepackage{bm}

\begin{document}
\title{$J$- factors of short DNA molecules}

\author{Marco Zoli}

\affiliation{School of Science and Technology \\  University of Camerino, I-62032 Camerino, Italy \\ marco.zoli@unicam.it}

\date{\today}

\begin{abstract}
The propensity of short DNA sequences to convert to the circular form is studied by a mesoscopic Hamiltonian method which incorporates both the bending of the molecule axis and the {intrinsic twist of the DNA strands}.  The base pair fluctuations with respect to the helix diameter are treated as path trajectories in the imaginary time path integral formalism. The partition function for the sub-ensemble of closed molecules is computed by imposing chain ends boundary conditions both on the radial fluctuations and on the angular degrees of freedom.
The cyclization probability, the $J$-factor, proves to be highly sensitive to the stacking potential, mostly to its nonlinear parameters. We find that the $J$-factor generally decreases by reducing the sequence length ($N$) and, more significantly, below $N  =\,100$ base pairs. However, even for very small molecules, the $J$-factors remain sizeable in line with recent experimental indications. Large bending angles between adjacent base pairs and anharmonic stacking appear as the causes of the helix flexibility at short length scales.
\end{abstract}


\maketitle

\section*{I. Introduction}

The probability for polymer chains to close into a ring, a long standing issue in physical chemistry, is traditionally addressed by the Jacobson-Stockmayer theory \cite{jacob}
which defines the propensity for cyclization i.e., the $J$- factor, as the ratio of equilibrium constants for unimolecular ring formation and bimolecular association (dimerization rate) \cite{flory}. Ring closure probabilities of DNA sequences have been extensively analyzed since thirty five years both experimentally \cite{shore} and theoretically \cite{olson}, as this method provides a quantitative measure of the helix twisting and bending flexibility which is key to the DNA packaging in chromosomes and to a variety of cellular processes. 

{In eukaryotic chromosomes DNA coils tightly around a histone octamer forming a nucleosome, the basic unit of chromatine \cite{bates}. This first level of compaction involves a stretch of about $147$ base pairs. Bacteriophages, widely used in genetic engineering, use proteins as molecular motors to condense their genomes (of various sizes)  inside pre-formed capsids whose diameters can be as small as $\sim 43 \,nm$ in phage RRH1 \cite{petrov}. Then, the functioning of DNA-packaging machines
in living organisms requires knowledge of the genome flexibility at scales of order of the typical persistence length or even shorter.
}

Given an ensemble of open ends chains, closure probabilities are governed by the competition of enthalpic effects which discourage the loop formation for \textit{short} molecules and entropic effects which inhibit loop formation in \textit{long} molecules with a large conformational space available for the open configurations:
due to thermal fluctuations, the two end-points of the chain are unlikely to come into contact.
Consistently, the experimental $J$- factor \cite{shore} shows a non-monotonous behavior versus the number ($N$) of base pairs (\textit{bps}) in the chain, smoothly decreasing above the peak located at $N \sim 500$, i.e., about three times the typical DNA persistence length. On the other side, below $N \sim 500$, the $J$- factor decreases \cite{shore1} with oscillations whose period provides a measure of the number of \textit{bps} per helix turn  \cite{note1}. 

While these findings had been interpreted both analytically by continuous worm-like chain (WLC) models \cite{shimada} and, numerically, by Monte Carlo simulations \cite{levene} and Hamiltonian methods accounting for the discreteness of the DNA molecules \cite{crothers1}, a renewed interest on the subject was raised some ten years ago when it was reported \cite{widom} that ligase assisted cyclization could occur in very short sequences, $N \sim 100$, with $J$- factors much higher than those predicted, albeit not previously measured, by conventional models. This report was soon after questioned  \cite{volo05} in view of the high ligase concentration used in the cyclization assay. Moreover, cyclization causes a high bending stress which enhances the probability of kinks at the sites of single stranded breaks \cite{kame04}: this, in turn, may alter the distributions of DNA fragments with joined (unligated) sticky ends in the circles with respect to the dimers thus invalidating a key assumption for the application of the ligation experimental method to the determination of the $J$- factor in very short molecules \cite{peters}. More recently however, a cyclization assay for single molecules which does not depend on external enzymes, the fluorescence resonance energy transfer (FRET), has yielded high looping rates for molecules with $N \sim 100$ \cite{vafa} supporting the 
conclusions of ref.\cite{widom} regarding the breakdown of the WLC model in the short lengths limit. High $J$- factors have also been reported by a similar FRET-based assay of the looping times of double stranded DNA \cite{kim13} whereas a careful examination of the unlooping rates as a function of the loop size \cite{kim14} has suggested that the anomalous $J$- factors may be reconciled with WLC analysis provided that the latter allow for kink formation in the strong bending regime.

The view that DNA flexibility exists at scales shorter than the persistence length has also been corroborated by measurements of end-to-end distance distributions based on x-ray scattering interference \cite{fenn} although the precise length scale for the likelihood of highly bent DNA conformations is still matter of debate \cite{forties,mazur,tan}.

On the theoretical side, it was predicted long ago \cite{crick} that the intrinsic bendability of the double helix, not necessarily a short one, may be due to kinks which maintain the base pairing but bring one base pair out of the stack, locally reducing the bending energy. In fact, it has later been shown that kinks can increase the cyclization efficiency in short sequences \cite{volo05}, but similar results could be also ascribed to bubble formation associated to the breaking of a few \textit{bps}  \cite{yan04} although this event has a higher (than the kink) energetic cost. 
Interestingly, single-strand-specific endonucleases applied to mini-circles with various $N \sim (65 - 105)$ have suggested  \cite{volo08} that  kinks in the double helix may indeed exist in the short sequences ($N \sim 65$) but the question of a critical molecule size for the appearance of helical disruptions still lacks a thorough understanding.

In some recent papers \cite{io14,io15},  the thermodynamic stability of a set of double stranded mini-circles has been investigated via path integral techniques and, by computation of the free energy, the stablest helicoidal conformations have been selected as a function of $N$. 
Mostly, it has been found that, for $N < 100$, the helix unwinding is inversely proportional to the molecule size.  Then, the mini-circles reveal a general tendency to untwist in order to release the stress incorporated in the highly bent circular conformations whereas thermal fluctuational effects may transiently drive the ensemble from the free energy minimum to one of the energetically close states. It is important to realize that, given a circle size with fixed $N$,  {the helical repeat $h$ (that is, the number of base pairs per helix turn)} has been determined as the most probable value for a large ensemble of molecule conformations participating to the partition function. As a main assumption, in Ref.\cite{io14}, the bending angle between adjacent base pair planes was kept constant thus neglecting those bending fluctuations in the structure of the stacking potential.

In this paper we address the issue of the  DNA flexibility by a different viewpoint, focusing on the open topic of the $J$-factor in short chains.  Specifically, a more general model is developed to include also the bending fluctuations among adjacent nucleotides stacked along the axis of the molecules. A broad range of bending angles is assumed in the open ends conformation and, by imposing appropriate closure conditions, the cyclization propensity of the molecules is computed as a function of $N$. Here $h$ is set to a constant such that $N / h$ is always an integer for our ensemble of molecules and no extra twist is necessary to close the chain into a loop. Thus, the cyclization  process does not require the unwinding of the complementary strands, namely it occurs at fixed helical repeat. 
It is emphasized that this constraint is not intrinsic to the model which, instead, can be formulated by thinking of $h$ as a system variable to be determined by free energy minimization criteria as e.g., in ref.\cite{io14}.

While this study assumes that, once formed, the circular molecules are in the double stranded helicoidal form,  it should be remarked that such assumption may not be appropriate for very small $N$ as only single stranded helices may be flexible enough to exist in such limit. However, the threshold below which circular helices can only be single stranded has not been settled yet and different groups have in fact detected (or built) double stranded mini-circles even in the $N \sim (60 - 80)$ range \cite{volo08,dutta}. Accordingly, these are the lengths of the shortest molecules hereafter considered. 
{The helical model is presented in Section II while the Hamiltonian model for a molecular chain is discussed in Section III. Section IV outlines the path integral method which is used to compute, in Section V, the free energy dependence on the model parameters. The $J$- factor analysis is contained in Section VI and some final remarks are given in Section VII.
}

\section*{II. Model for the Helix}

\begin{figure}
\includegraphics[height=8.0cm,width=8.0cm,angle=-90]{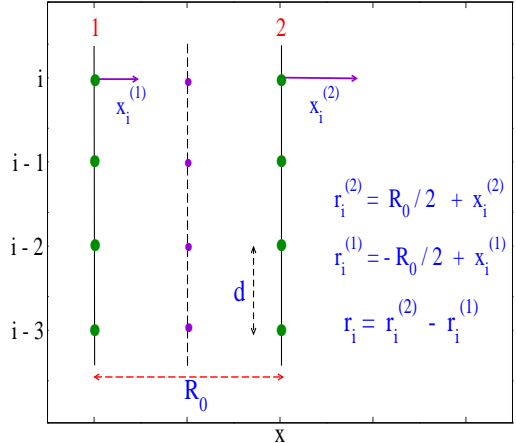}
\caption{\label{fig:1}(Color online)  
Schematic of the model for $N$ base pairs.  $N$ bases (green dots) are stacked along each of the two complementary strands in a ladder representation. $R_0$ is the equilibrium inter-strand distance and $d$ is the rise distance. $x_{i}^{(1,2)}$ are the fluctuations of the $i-th$ base pair mates with respect to the equilibrium.
$r_{i}$ is the relative distance between the two mates. }
\end{figure}

To begin with, we adopt a general  picture for a double stranded chain whose \textit{bps} are modeled as shown in Fig.~\ref{fig:1}.

In this simple ladder representation, the two mates of the $i-th$ base pair can fluctuate around their equilibrium positions represented by the green dots lying along the two complementary strands. 
The vibrations of the two bases along the stack are much smaller than the transverse vibrations $x_{i}^{(1,2)}$, i.e. the model is at this stage one-dimensional. $x_{i}^{(1)}$ and $x_{i}^{(2)}$ may be in-phase (as depicted) or out-of-phase. In general, also their amplitudes may differ. 
$R_0=\,20 $ \AA {} is the average helix diameter and  $d=\, 3.4$ \AA {} is the average rise distance. 
With respect to the central helical axis (that is kept fixed), we build the vectors $r_{i}^{(1)}=\, -R_0 / 2 + x_{i}^{(1)}$ and $r_{i}^{(2)}=\, R_0 /2 + x_{i}^{(2)}$ and define the relative distance $r_{i}=\, r_{i}^{(2)} - r_{i}^{(1)}$ which will be the object of our path integral analysis. Note that: \textit{i)} also in-phase vibrations of different amplitudes may contribute to $r_{i}$ shifting the base pair out of the stack.  \textit{ii)} $r_{i}$ may shrink with respect to $R_0$ but too large contractions are prevented by the strands electrostatic repulsion.

\begin{figure}
\includegraphics[height=8.0cm,width=8.0cm,angle=-90]{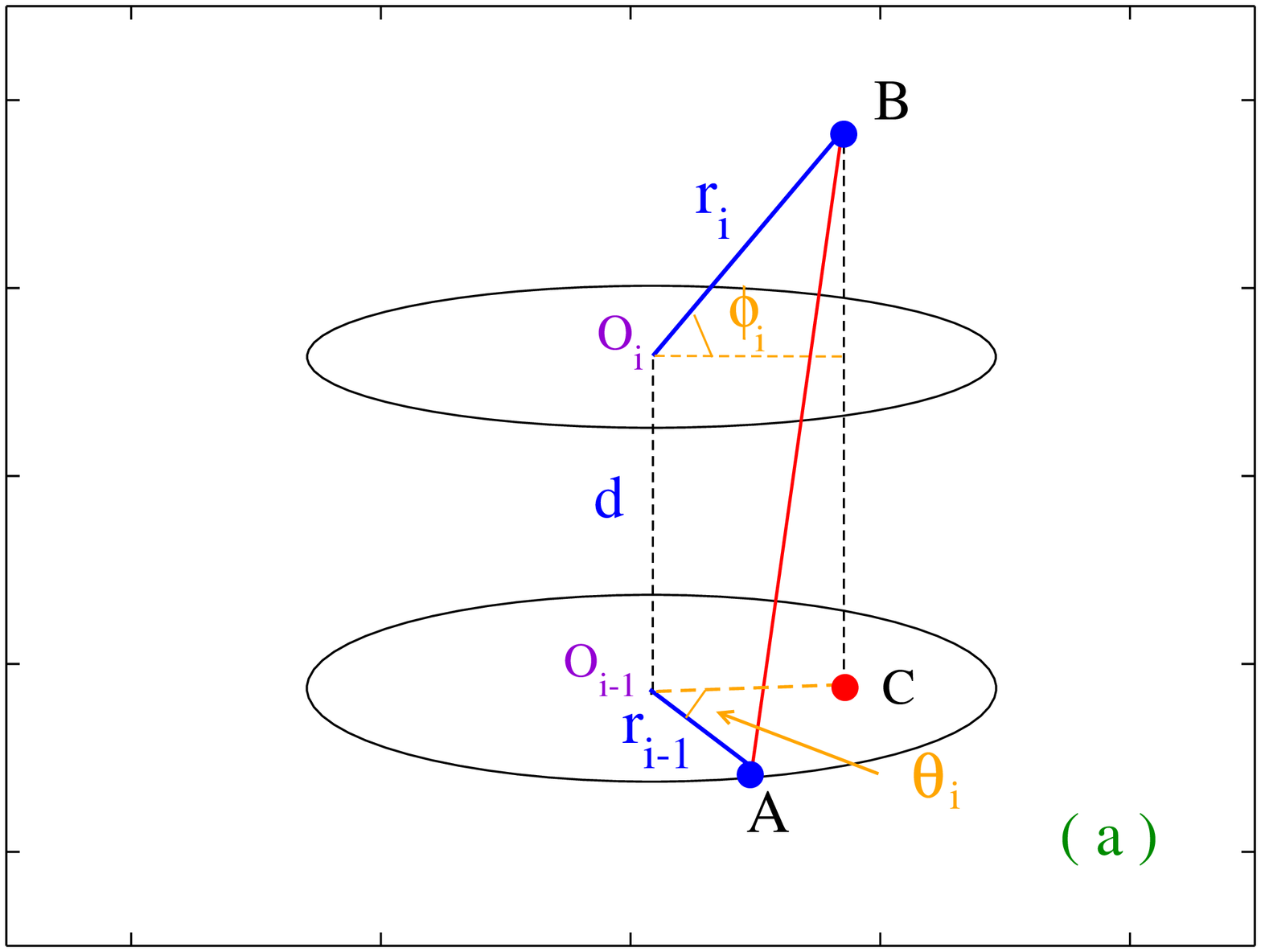}
\includegraphics[height=8.0cm,width=8.0cm,angle=-90]{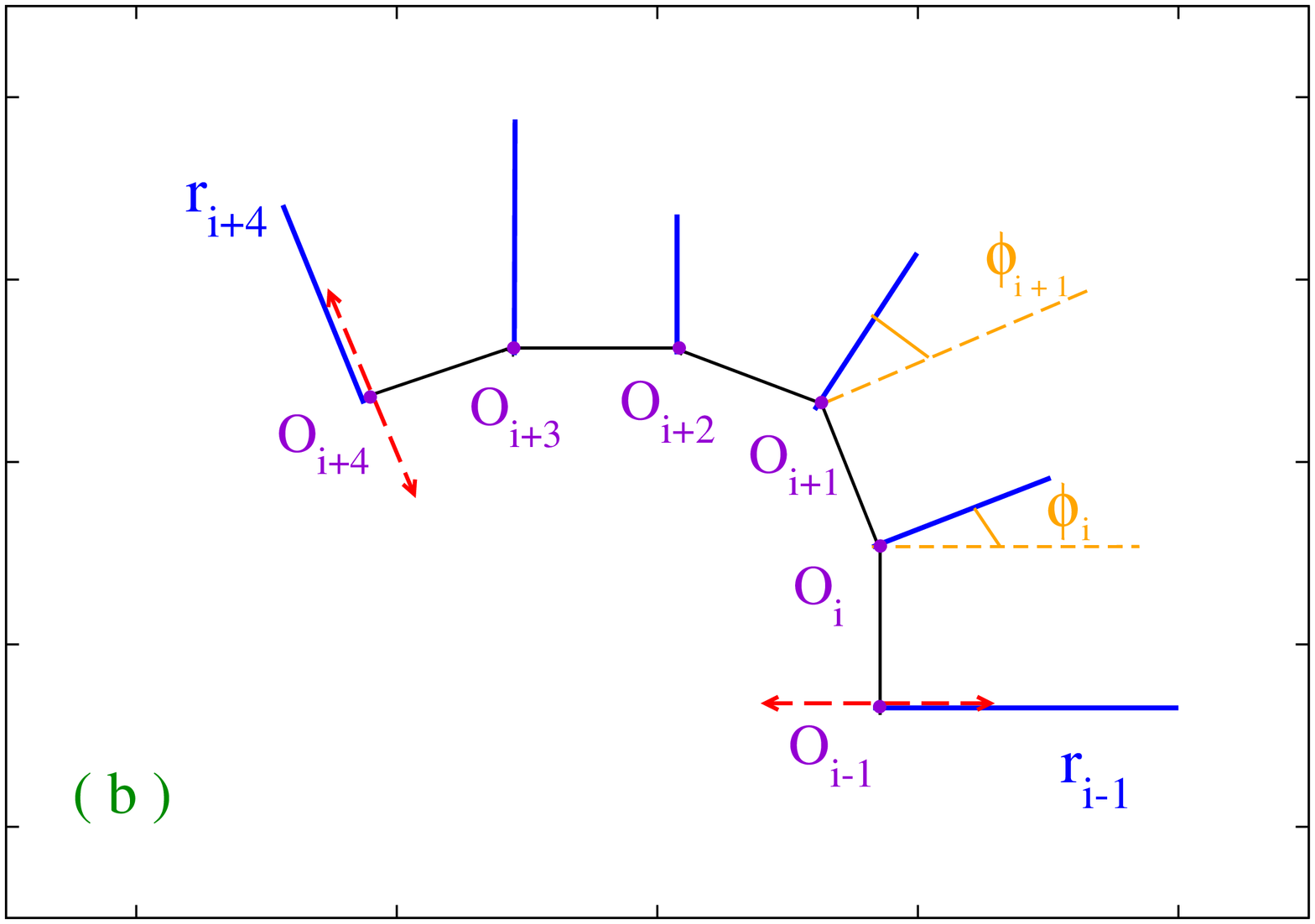}
\caption{\label{fig:2}(Color online)  
(a) Beyond the ladder representation: $\theta_i$ is the twist angle between adjacent relative distances, $r_{i-1}$ and $r_i$. The $O_i$'s correspond to the violet dots in Fig.~\ref{fig:1} lying along the central helical axis. The ovals depict the fact that base pair vectors can take arbitrary orientations (selected by $\theta_S$ in the text) in the plane normal to the sheet. {} $\phi_{i}$ is the (variable) bending angle between adjacent base pair vectors.  (b) The effect of the bending is further visualized by setting, $\theta_i=\,0$.  The helix axis is a chain of $N - 1$ segments of length $d$ connecting the $O_i$'s. As the latter are pinned to the sheet plane, the helix axis is planar hence, for the closed chain conformation, the writhe is zero. The (red) long-dashed lines, drawn at the $O_{i-1}$ and $O_{i+4}$ sites, denote the helix diameter $R_0$.
The $r_i$'s depart from the $O_i$'s, have variable amplitudes and are parallel to $R_0$ at their respective sites. The $\phi_{i}$'s are measured from the (orange) short-dashed lines which are parallel to the adjacent (preceding) $r_{i-1}$'s along the chain.
}
\end{figure}

Some coarse assumptions are inherent to this mesoscopic modeling: it does not contain the spatial extension of the bases which instead are point-like objects overlapping with their respective sugar-phosphate groups. Accordingly, the Hamiltonian contains effective parameters accounting for the inter- and intra-strand interactions between nucleotides. Moreover, only the relative distance between the pair mates is considered in the work whereas other degrees of freedom, such as propeller twisted base pairs distorting the hydrogen bonds, are not.

Next, we go beyond the ladder model and admit that $r_{i}$ can take any orientation in a plane normal to the sheet plane: say $\theta_i$ the angle between adjacent 
{$r_{i-1}$} and {$r_{i}$} along the stack, see Fig.~\ref{fig:2}(a). This amounts to introduce a twist between \textit{bps} in a fixed planes representation  \cite{io11}. For the $i-th$ base pair, the torsional angle is given by, $\theta_i =\, (i - 1)\theta  + \theta_S$,  where $\theta=\, 2\pi / h$  and $h$ is the number of \textit{bps} per helix turn. Throughout this work we take, $\,h=\,10$, about the standard value measured for instance in covalently closed DNA in solution \cite{wang}. $\theta_S$ is the twist of the first base pair along the stack. To keep the model general, we sum over a distribution of $\theta_S$'s thus fulfilling the above mentioned requisite of general base pair orientation. 
Furthermore, we release the \textit{fixed planes} constraint and, instead, admit that adjacent planes may be inclined by the angle $\phi_i$ as shown in Fig.~\ref{fig:2}(a): bending fluctuational effects are introduced by taking $\phi_i$ as an integration variable. 
Accordingly, the square distance between {$r_{i-1}$} and {$r_{i}$} (measured from $O_{i-1}$ and $O_i$, respectively) is:

\begin{eqnarray}
& & \overline{AB}^2=\, \overline{BC}^2 + \overline{AC}^2 \, , \nonumber
\\
& & |\overline{BC}| =\, d + r_i \sin \phi_i  \, , \nonumber
\\
& & \overline{AC}^2=\, r_{i-1}^2 + \overline{O_{i-1}C}^2 -2 r_{i-1} \cdot |\overline{O_{i-1}C}| \cos \theta_i \, , \nonumber
\\
& & |\overline{O_{i-1}C}| =\, r_i \cos \phi_i  \, .
\label{eq:0}
\end{eqnarray}

{Eq.~(\ref{eq:0}) is used in the next Section to represent the stacking interactions. 

The role of the bending in this model is further elucidated by Fig.~\ref{fig:2}(b), where the torsional angle is suppressed and the chain, together with the $r_{i}$'s real space trajectories,  is drawn in two-dimensions. While, by construction, the $r_{i}$'s are at any site parallel to $R_0$, their variable amplitudes characterize a specific molecule configuration according to the method described in Section IV. It follows that the helix diameters at adjacent sites, $O_{i-1}$ and $O_i$, are also bent by the variable  $\phi_i$ while the molecular axis is a chain of $N-1$ segments lying on a plane. Importantly, this latter property does not depend on the 2D drawing reported in  Fig.~\ref{fig:2}(b). The $O_i$'s are always pinned to the sheet plane also in the presence of the torsional degree of freedom. Hence, whenever the conditions for the formation of a circular chain are fulfilled, the writhe of that circular molecule is zero.
This is consistent with the shortness of the molecules hereafter considered  \cite{bates,lionb,irob}.

{Certainly, in open ends sequences, the bending of the molecular axis should not be confined to a single plane and a broader ensemble of fluctuations may favor a higher conformational entropy. This effect may somewhat yield a larger contribution to the partition function of the open ends configuration than that we are estimating in this study. Nevertheless such effect is expected to be more significant in longer sequences.
}

\section*{III. Hamiltonian Model }

The fundamental interactions at play in the open ends heterogeneous chain with $N$ nucleotides, depicted in Fig.~\ref{fig:2}, are represented by the following mesoscopic Hamiltonian:

\begin{eqnarray}
& &H =\, H_a[r_1] + \sum_{i=2}^{N} H_b[r_i, r_{i-1}] \, , \nonumber
\\
& &H_a[r_1] =\, \frac{\mu}{2} \dot{r}_1^2 + V_{1}[r_1]  \, , \nonumber
\\
& &H_b[r_i, r_{i-1}]= \,  \frac{\mu}{2} \dot{r}_i^2 + V_{1}[r_i] + V_{2}[ r_i, r_{i-1}, \phi_i, \theta_i]   \, , \nonumber
\\
& &V_{1}[r_i]=\, V_{M}[r_i] + V_{Sol}[r_i] \, , \nonumber
\\
& &V_{M}[r_i]=\, D_i \bigl[\exp(-b_i (|r_i| - R_0)) - 1 \bigr]^2  \, , \nonumber
\\
& &V_{Sol}[r_i]=\, - D_i f_s \bigl(\tanh((|r_i| - R_0)/ l_s) - 1 \bigr) \, , \nonumber
\\
& &V_{2}[ r_i, r_{i-1}, \phi_i, \theta_i]=\, K_S \cdot \bigl(1 + G_{i, i-1}\bigr) \cdot \overline{AB}^2  \, , \nonumber
\\
& &G_{i, i-1}= \, \rho_{i, i-1}\exp\bigl[-\alpha_{i, i-1}(|r_i| + |r_{i-1}| - 2R_0)\bigr]  \, . \nonumber
\\ 
\label{eq:02}
\end{eqnarray}

{ Each base pair (except the two end sites) interacts with its two adjacent neighbors, see Fig.~\ref{fig:2}(b). As only the first site, $i=\,1$, lacks the preceding base pair along the chain,  its kinetic term and one particle potential have been treated separately by defining $H_a[r_1]$.
Note also that the first site is coupled to the second one via the $i=\,2$ term in $H_b[r_i, r_{i-1}]$. }

By the Hamiltonian in Eq.~(\ref{eq:02}) we propose a general description for an open ends molecule with finite helical radius and base pairs stacked along the molecular axis according to a specific intra-strand potential. The relative base distances are measured with respect to the helix diameter that sets the zero for the potential. 
The implementation of this feature in the computational method is discussed in Section IV.

The one particle potential, $V_{1}[r_i]$, includes two contributions: 

\textit{(a)} the Morse potential $V_{M}[r_i]$ modeling the hydrogen bond stretching vibrations between complementary bases: $D_i$ is the pair dissociation energy and $b_i$ determines the potential range. 

Fluctuations in the base pair separations may reduce the distance $r_i$ between complementary strands to values smaller than $R_0$, a case also contemplated at some sites in Fig.~\ref{fig:2}(b).
However, such reduction is limited by the hard core electrostatic repulsion due to the negatively charged phosphate groups. To comply with this physical requirement, the numerical code discards $r_i$ such that, $|r_i| - R_0 < - \ln 2 / b_i$ which would deliver a repulsive energy larger than $D_i$. This sets the link between the fundamental parameters of our analysis. Adenine-Thymine \textit{bps} can be broken more easily and undergo larger stretching vibrations than Guanine-Cytosine \textit{bps} \cite{singh,kalos}. Thus we set: $D_{AT} < D_{GC}$ and $b_{AT} < b_{GC}$ noticing that, with regard to their effective values, substantial variations have been reported depending both on the model and on the sequence properties \cite{campa,weber09,weber15}.  Here, $D_i$ and $b_i$ are tuned in order to yield a free energy per base pair in line with the experimental data \cite{kame06}.

\textit{(b)} The term $V_{Sol}[r_i]$ accounting for the fact that DNA is always immersed in water \cite{zanc}. Then,  the molecules stability depends on the counter-ion concentration in the solvent {which can be empirically related to the $f_s$ parameter \cite{druk}. As a main effect, the solvent potential enhances by $f_s D_i$ (with respect to the Morse plateau) the height of the energy barrier above which the base pair dissociates. Thus, the full one particle potential, $V_{M}[r_i] + V_{Sol}[r_i]$, shows a hump whose width is tuned by $l_s$. This length defines the range within which $V_{Sol}$ is superimposed to the plateau of the Morse potential.   }

While the solvent term has been discussed in ref.\cite{io12}, its parameters are taken constant hereafter by setting $f_s=\,0.1$ and $l_s=\,0.5$ \AA.  
An extensive analysis of the interplay between salt concentration and potential parameters which control the stability of heterogeneous DNA molecules can be found in refs.\cite{weber15,singh15}  with regard to mesoscopic models for the helix.

DNA cyclization crucially depends on the flexibility of the molecule backbone \cite{desantis,croth00}. Accordingly we model the stacking by a two particles potential, $V_{2}[ r_i, r_{i-1}, \phi_i, \theta_i ]$, containing both the twisting of the helix and the bending fluctuations. The square distance between adjacent $r_i$ and $r_{i-1}$, that is $\overline{AB}^2 $,  is given in  Eq.~(\ref{eq:0}).

The nonlinear potential was originally proposed \cite{pey2}, in the context of the thermally {driven denaturation of a ladder DNA model}, to describe those cooperative effects which propagate along the molecule stack forming large base pair openings at high temperature. The nonlinear features have been maintained in the stacking potential {which has been here generalized to the more structured helical model of} Figs.~\ref{fig:2}. The underlying idea is that, whenever $r_{i}  - R_0 \gg \alpha_{i, i-1}^{-1}$, the $i-th$ hydrogen bond is broken and the stacking coupling drops from \, $\sim K_S \cdot (1 + \rho_{i, i-1})$ to $\sim K_S$:  this also favors the breaking of the adjacent base pair and the consequent opening of local bubbles \cite{cule,benham,zocchi03,bonnet,ares,rapti,metz12,palmeri13}. 

In this regard, our stacking potential is more complex than the usual elastic terms assumed in WLC models \cite{volo2000} and also in sequence dependent simulations of DNA configurations accounting for the base-pair steps \cite{olson06}.

Then, the $\alpha_{i, i-1}$ measure how large the opening of a base pair should be to produce such a reduction in the stacking. If the condition \,\,  $\alpha_{i, i-1} < b_i$ \,\, is fulfilled, the range of the stacking is larger than that of the Morse potential and large fluctuations are required to unstack a base pair. 
A weak harmonic stiffness constant, $K_S=\,10 \, meV \AA^{-2}$, is assumed while the effects of nonlinear path displacements on the cyclization probability are discussed by tuning the parameters $\alpha_{i, i-1}$ and $\rho_{i, i-1}$. 

We feel that the specific $V_{2}$ in Eq.~(\ref{eq:02}) has robust physical motivation, although it should be pointed out that such choice is not unique: in fact different potentials may be taken e.g., with the purpose to ensure the finiteness of the intra-strand stacking also for large inter-strand separation \cite{joy09}. This requirement is fulfilled in our computational method by truncating the phase space available to the base pair separations. 
  
Importantly, this technique has the advantage to tackle the problem of the divergence of the partition function for the Hamiltonian in Eq.~(\ref{eq:02}), encountered e.g., in transfer integral techniques \cite{zhang}. Such problem arises from the fact that the one-particle potential is bounded for $r_i \rightarrow \infty$. Then, if all $r_i$'s are equal (translational mode) and infinitely large, the two-particles potential vanishes while $H$ remains finite hence the partition function diverges. This zero mode cannot be removed via standard techniques \cite{schulman} due to the lack of translational invariance caused by the on-site potential.   Further details of the computational method are given in ref.\cite{io14}.

\section*{IV. Method}

The heterogeneous system of $N$ purine-pyrimidine \textit{bps} with reduced mass $\mu$, given in Eq.~(\ref{eq:02}),  is treated in the finite temperature path integral formalism widely presented in the last years \cite{io11}.
The motivations and key features of the method are summarized hereafter.

Essentially, the one dimensional base pair displacements are mapped onto the time axis, $r_i \rightarrow |r_i(\tau)|$,  so that the distance between the base pair mates
is a trajectory depending on the imaginary time  $\tau=\,it$,  with $t$ being the real time for the path evolution amplitude within the time interval, $t_b - t_a$. The theoretical grounds of the method lie in the analytic continuation of the quantum mechanical partition function to the imaginary time axis which, in general, permits to get the quantum statistical partition function \cite{fehi}. Accordingly $\tau$ varies in a range $\tau_b - \tau_a$ whose amplitude is set by the inverse temperature $\beta$ \cite{feyn} and the partition function is written as an integral over closed trajectories running along the $\tau$-axis. 

While the imaginary time formalism  is widely used in semi-classical methods for the solution of quantum statistical problems \cite{jack},  
our method extends the  $\tau$-formalism to the classical regime, the appropriate one to treat the room temperature DNA molecules. 
This is done by applying the same formal replacement which permits to solve exactly
the partition function of the harmonic Hamiltonian for the ladder model in Fig.~\ref{fig:1} \cite{pey04}. For the latter, the solution is found by mapping the associated transfer integral equation onto a Schr\"{o}dinger equation for a quantum particle in a Morse potential as described in detail in ref. \cite{io14b}. 
         
As a consequence of the $\tau$-closure condition  $(\,r_i(0)=\, r_i(\beta) \,)$,  the $r_i(\tau)$ can be written in Fourier series:

\begin{eqnarray}
& &r_i(\tau)=\, R_0 + \sum_{m=1}^{\infty}\Bigl[(a_m)_i \cos(\omega_m \tau ) + (b_m)_i \sin(\omega_m \tau ) \Bigr] \, , \nonumber
\\
& &\omega_m =\, \frac{2 m \pi}{\beta} 
\label{eq:01}
\end{eqnarray}

and this expansion defines the associated integration measure $\oint {D}r_i$  over the space of the Fourier coefficients:

\begin{eqnarray}
& &\oint {D}r_{i} \equiv  \prod_{m=1}^{\infty}\Bigl( \frac{m \pi}{\lambda_{cl}} \Bigr)^2 \int_{-\Lambda_T}^{\Lambda_T} d(a_m)_i \int_{-\Lambda_T}^{\Lambda_T} d(b_m)_i \, , \, \nonumber
\\
\label{eq:04}
\end{eqnarray}

where $\Lambda_T$ is the temperature dependent cutoff.  $\lambda_{cl}$ is the classical thermal wavelength which depends on $K_S$ as shown in ref.\cite{io14b}. As $K_S$ is kept constant, also $\lambda_{cl}$ is a constant parameter of the model.

The distinctive features of the imaginary time path integral formalism are:

{\textit{i)}} By mapping the real time derivative onto the imaginary time derivative, $\frac{d}{dt} \rightarrow  \, i \frac{d}{d\tau } $, one introduces a sign change in the imaginary time kinetic action with respect to the real time kinetic action. Accordingly, the partition function is obtained  by computing the fluctuational effects (associated to the Fourier coefficients in Eq.~(\ref{eq:01})) around the classical path, $r_i(\tau) \sim \, R_0$, which minimizes the \textit{sum} of the kinetic and potential term in the Euclidean action.

{\textit{ii)}}
Intrinsic to the path integration technique \cite{io05} is the condition that the measure in Eq.~(\ref{eq:04}) normalizes the kinetic term in the action, i.e.:

\begin{eqnarray}
\oint {D}r_i \exp\Bigl[- \int_0^\beta d\tau {\mu \over 2}\dot{r}_i(\tau)^2  \Bigr] = \,1 \, .
\label{eq:11} \,
\end{eqnarray}

This condition consistently defines the cutoffs $\Lambda_T$ in the path phase space  \cite{io11a} avoiding those indeterminacies peculiar of the transfer integral methods \cite{erp}.
Also note that Eq.~(\ref{eq:11}) holds for any $\mu$. This amounts to say that the system free energy does not depend on $\mu$, as expected for a classical system.
Moreover, the measure in Eq.~(\ref{eq:04}) permits to integrate both kinetic and potential actions over the same degrees of freedom thus avoiding the decoupling between momenta and real space integrations operated in the usual approach to the classical partition function, see e.g. ref.\cite{pey04}. Accordingly, Eq.~(\ref{eq:04}) correctly renders a dimensionless total partition function.

{\textit{iii)}} Eq.~(\ref{eq:01})  generates a large ensemble of path amplitudes for any base pair. Say \, $2N_p + 1$ \, the number of integration points for each Fourier coefficient  in Eq.~(\ref{eq:04}). Then, for a single Fourier component, the computation includes \, $(2N_p + 1)^2 \cdot N_\tau $ \, paths where $N_\tau $ is the number of points in the imaginary time integration. This is the base pair ensemble size in the path phase space. The number of paths is increased until numerical convergence in the partition function is achieved. These paths have to fulfill the physical requirements described in Section III. Thus our numerical program selects, at any $T$, an ensemble of \textit{good paths} which are \textit{1)} consistent with the model potential constraints and \textit{2)} in accordance with the second law of thermodynamics \cite{io11}.

Note however that there is a significant difference between Eq.~(\ref{eq:01}) and the Fourier series representation of the base pair displacements used in previous papers, e.g., ref.\cite{io14}: this deserves some discussion.

It has been mentioned above and shown in Fig.~\ref{fig:1}, that the base pair separations in the current model are measured with respect to the helix diameter which, instead, had not been defined in ref.\cite{io14}.
Accordingly, in the current path integral description, the path amplitudes should fluctuate with respect to $R_0$. This is accomplished, in Eq.~(\ref{eq:01}), by setting the usual zero mode $(r_0)_i$ equal to $R_0$. Consistently, the integration measure in Eq.~(\ref{eq:04}) does not contain the $\int d(r_0)_i$.   

Certainly one might have maintained the standard Fourier expansion (with the
$(r_0)_i$ term) also in the present calculation but, in this case, one should have varied the coefficients $\{(r_0)_i, (a_m)_i, \, (b_m)_i\}$ within a much larger path configuration space than that required by Eq.~(\ref{eq:01}). Note in fact that the paths mostly contributing to the partition function are those which minimize the action, namely the paths such as \, $r_{i}(\tau) - R_0$ \, is a small quantity. This means that, applying the standard expansion rather than Eq.~(\ref{eq:01}), one should have taken  larger cutoffs in the integration measure thus building a much larger paths ensemble (at the price of a much longer CPU time) and, eventually, using only a sub-ensemble in the path integration while discarding all those paths whose amplitude is much smaller than $R_0$. 

Also in view of the fact that the two options, standard expansion and Eq.~(\ref{eq:01}), have proved to yield similar results for the partition function, I have followed the latter option in the calculations presented in the next Sections.

Furthermore, Eq.~(\ref{eq:01}) is also fully consistent with the normalization condition in Eq.~(\ref{eq:11}) as the kinetic energy term
depends (in any case) only on the $\{(a_m)_i, \, (b_m)_i\}$ coefficients.

The analysis presented so far and markedly Eqs.~(\ref{eq:01}), ~(\ref{eq:04}), ~(\ref{eq:11}), make clear that our method is based on a one-dimensional path integration over the base pair separations. This method builds base pair paths whose amplitudes, mapped onto the $\tau$ axis, are growing function of temperature in agreement with general expectations and experimental data \cite{zocchi03,metz12}.
Instead, the angular variables, $\phi_i$ and $\theta_i$ in Eq.~(\ref{eq:01}), are treated in a conventional way. Specifically, as explained in Section II, a sum is performed over a distribution of twist angles at fixed helical repeat whereas a direct integration is carried out over a broad range of \textit{in-plane} bending fluctuations. While these approximations may be removed in a more general investigation, at the present stage there is not enough knowledge regarding the temperature dependence of the bending angles to justify a full (and time consuming) path integral approach such to incorporate a $T$-dependent cutoff on the bending fluctuations.

With these caveats, we can proceed to write the classical partition function, $Z_N$, in the path integral formulation. Consistently with the notation for the Hamiltonian  in Eq.~(\ref{eq:02}), $Z_N$ reads:

\begin{eqnarray}
& &Z_N=\, \oint Dr_{1} \exp \bigl[- A_a[r_1] \bigr]   \prod_{i=2}^{N} \sum_{\theta_S} \int_{- \phi_M }^{\phi_M } d \phi_i  \oint Dr_{i}  \exp \bigl[- A_b [r_i, r_{i-1}] \bigr] \, , \nonumber
\\
& &A_a[r_1]= \,  \int_{0}^{\beta} d\tau H_a[r_1(\tau)] \, , \nonumber
\\
& &A_b[r_i, r_{i-1}]= \,  \int_{0}^{\beta} d\tau H_b[r_i(\tau), r_{i-1}(\tau)] \, ,
\label{eq:03}
\end{eqnarray}

where the action $A_b[r_i, r_{i-1}]$ depends: \textit{(1)} on the Fourier coefficients 
$\{ (a_m)_i, \, (b_m)_i\}$ and $\{ (a_n)_{i-1}, \, (b_n)_{i-1}\}$
of the $i$ and $i-1$ base pair path amplitudes respectively; \textit{(2)} on the angles $\theta_S$ and $\phi_i$.
Thus, the two particle stacking potential brings about a mixing of the Fourier components of adjacent path amplitudes which largely enhances the computational time.

Kinks with even large bending angles are included in Eq.~(\ref{eq:03}) by taking a symmetric angular cutoff with \,  $\phi_M \sim \pi /2$ \cite{salari}.  This suffices to achieve numerical convergence: larger cutoffs would not add significantly changes to $Z_N$.

From Eqs.~(\ref{eq:02}),~(\ref{eq:03}), one notices that the largest contribution to $Z_N$ comes from those trajectories which minimize the sum of the kinetic and potential energy thus corroborating the above discussed choice for the path expansion.

\section*{V. Free energy}
 
The free energy per particle, $F_{N}=\, - (N \beta)^{-1} \ln Z_{N}$, is plotted in Figs.~\ref{fig:3} for a heterogeneous chain with $N=\,100$ and $50\%$ $AT$-\textit{bps}.

\begin{figure}
\includegraphics[height=8.0cm,width=8.0cm,angle=-90]{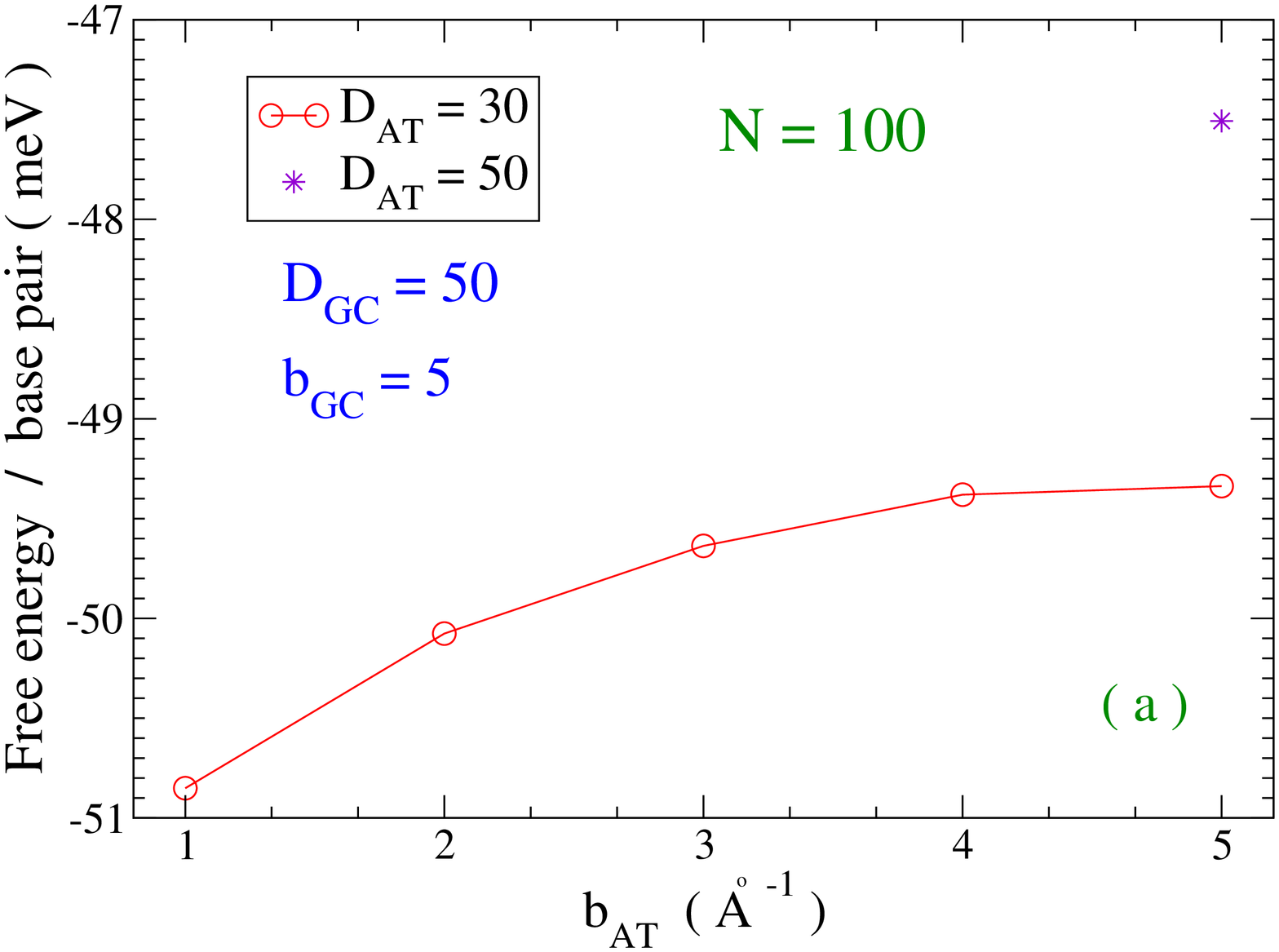}
\includegraphics[height=8.0cm,width=8.0cm,angle=-90]{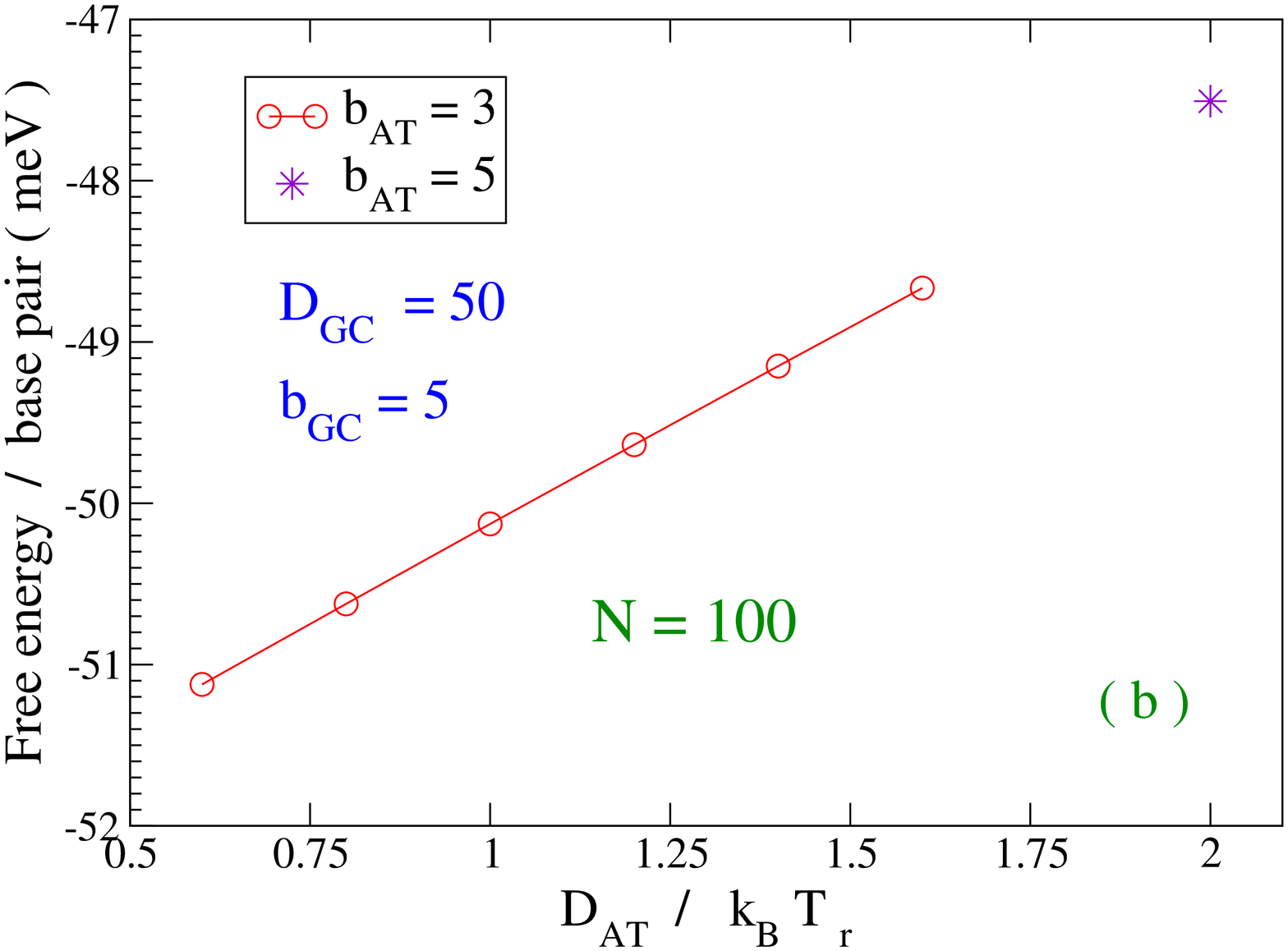}
\caption{\label{fig:3}(Color online)  
Free energy per base pair for an open ends chain molecule with $50 \%$ AT- base pairs.  (a) Free energy versus inverse length of the Morse potential for AT-\textit{bps}. $D_{AT}$ and $D_{GC}$ are in units $meV$ as in the text. (b) Free energy versus dissociation energy for AT-\textit{bps}. $D_{AT}$ is in units $k_B T_r = \, 25 \, meV$, that is the room temperature thermal energy. $D_{GC}$ is in units $meV$. Both in (a) and (b), the stars refer to the homogeneous chain with $100 \%$ GC-\textit{bps}.}
\end{figure}

In these calculations, heterogeneity is accounted for through the hydrogen bond Morse potential whereas the nonlinear stacking parameters are assumed to be homogeneous namely,  $\alpha_{i, i-1}\equiv \alpha _i = 2.5  \,\AA^{-1}$, $\rho_{i, i-1}\equiv \rho _i = 1$.  While the latter assumption may be consistently dropped in our model (see ref.\cite{io13} ) for analysis of specific sequences, it can be maintained in the present context in view of the strong parameter sensitivity (see next Section) displayed by the cyclization factors.

Note that $F_{N}$ computed via Eq.~(\ref{eq:03}) is an average value incorporating both the stacking and the inter-strand contributions whereas the two effects separately can yield significantly different stability parameters \cite{metz11}.
Keeping fixed the $GC$-parameters, we compute the free energy tuning $b_{AT}$ and $D_{AT}$.  As a general trend, $F_{N}$ decreases by increasing the amplitude of the path displacements (reducing $b_{AT}$ in Fig.~\ref{fig:3}(a)) and by lowering the energy threshold for pair breaking  (reducing $D_{AT}$ in Fig.~\ref{fig:3}(b)). The parameters  are such that $F_{N}$ is  $\, \sim 1 - 1.2 kcal/mol$, consistent with the experiments although large discrepancies exist among the data published by different groups \cite{santa}.  For comparison, also a homogeneous chain of $100$ $GC$-\textit{bps} is considered in Figs.~\ref{fig:3} (violet star symbol): in this case, $F_{N}$ is somewhat larger consistently with the expectation that this molecule has a lower conformational entropy as a consequence of the higher stability of the $GC$ bonds which reduce the overall flexibility.

\section*{VI. $J$- Factor}

After discussing the model dependence on the hydrogen bond parameters, we set out to calculate the $J$- factor that is the
ensemble probability of the circular conformation within a given capture volume: this varies, in our model, with the fundamental rise distance $d$. 

Formally, the cyclization probability is defined by:

\begin{eqnarray}
& &J=\, 8 \pi^2 \frac{Z_{cycle}}{Z_N}  \, , \nonumber
\\
& &Z_{cycle}=\,  \oint Dr_{1} \exp \bigl[- A_a[r_1] \bigr]  \prod_{i=2}^{N} \sum_{\theta_S} \int_{- \phi_M }^{\phi_M } d \phi_i  \oint Dr_{i}  \delta^3({\textbf{r} }_{i=\,1} - {\textbf{r} }_{i=\,N}) \exp \bigl[- A_b [r_i, r_{i-1}] \bigr] \bigr] \, . \nonumber
\\
\label{eq:05}
\end{eqnarray}

$Z_{cycle}$ is the partition function for the ensemble of molecules in a closed configuration and $Z_{N}$ is the general partition function in Eq.~(\ref{eq:03}) which lacks such constraint. 

{Only a fraction of molecule conformations contained in $Z_{N}$ align with a specific orientation. 
The factor $4 \pi \cdot 2 \pi$ in Eq.~(\ref{eq:05}) stems from the boundary constraints associated to the loop formation.
Precisely, the factor $4 \pi$ accounts for all possible loop orientations over solid angle and, once a specific orientation is set,
$2 \pi$ is the angular range for rotations around that orientation in order to achieve torsional alignment of the molecule end sites \cite{wilson,yan15}. 
}
Due to the three dimensional $\delta$-function, $J$  has the unit of an inverse volume that is one molecule over $d^3$. This is converted to $moles / liter$ dividing by the Avogadro's number which amounts to multiply $J$ in Eq.~(\ref{eq:05}) by the factor $42.26$. The boundary constraints expressed by the $\delta$-functions are implemented in the code by imposing that first and last base pairs in the sequence have, \textit{i)} same bending and twisting angles, \textit{ii)} same $r_i(\tau)$ (for any $\tau$). 

\begin{figure}
\includegraphics[height=8.0cm,width=8.0cm,angle=-90]{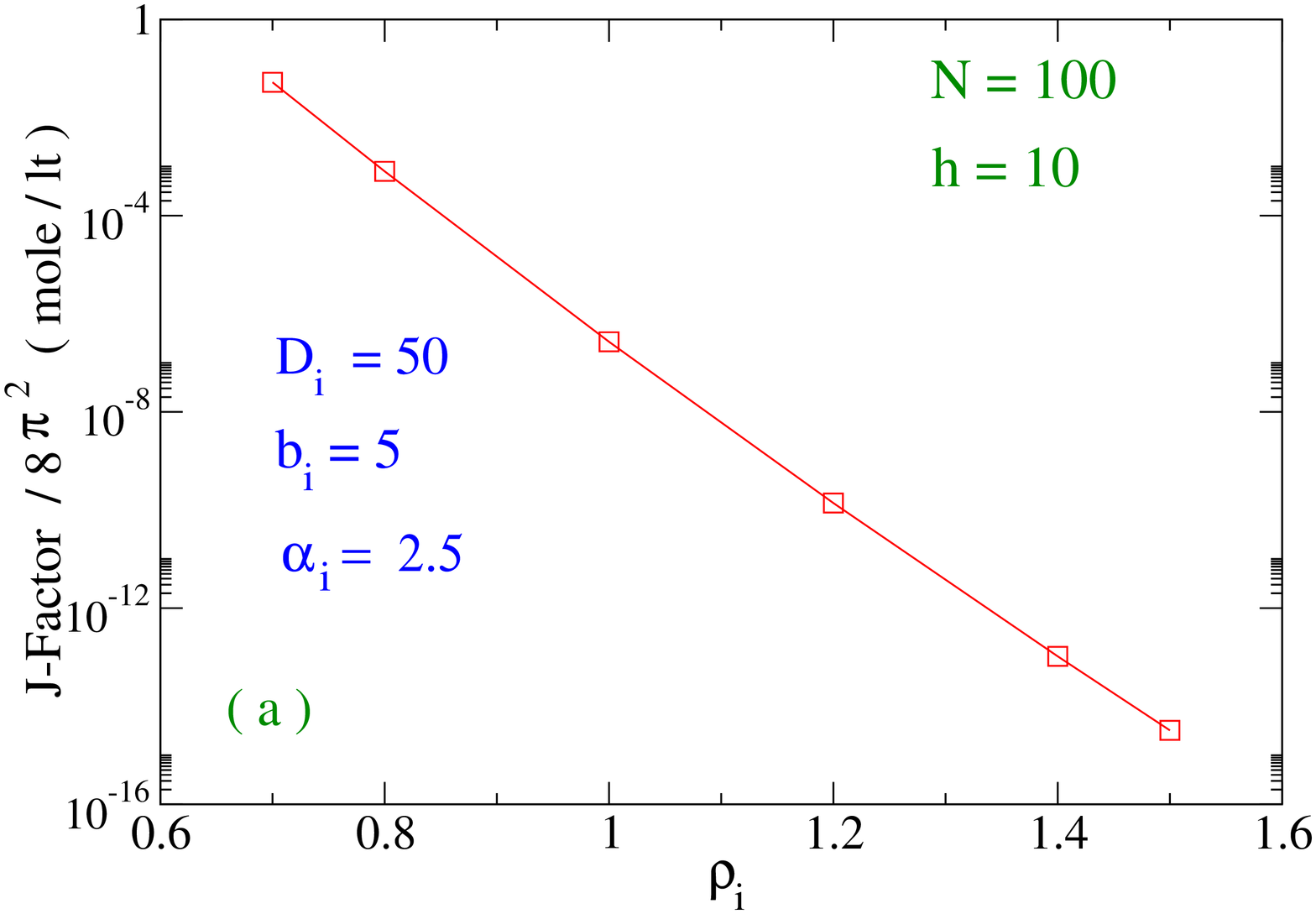}
\includegraphics[height=8.0cm,width=8.0cm,angle=-90]{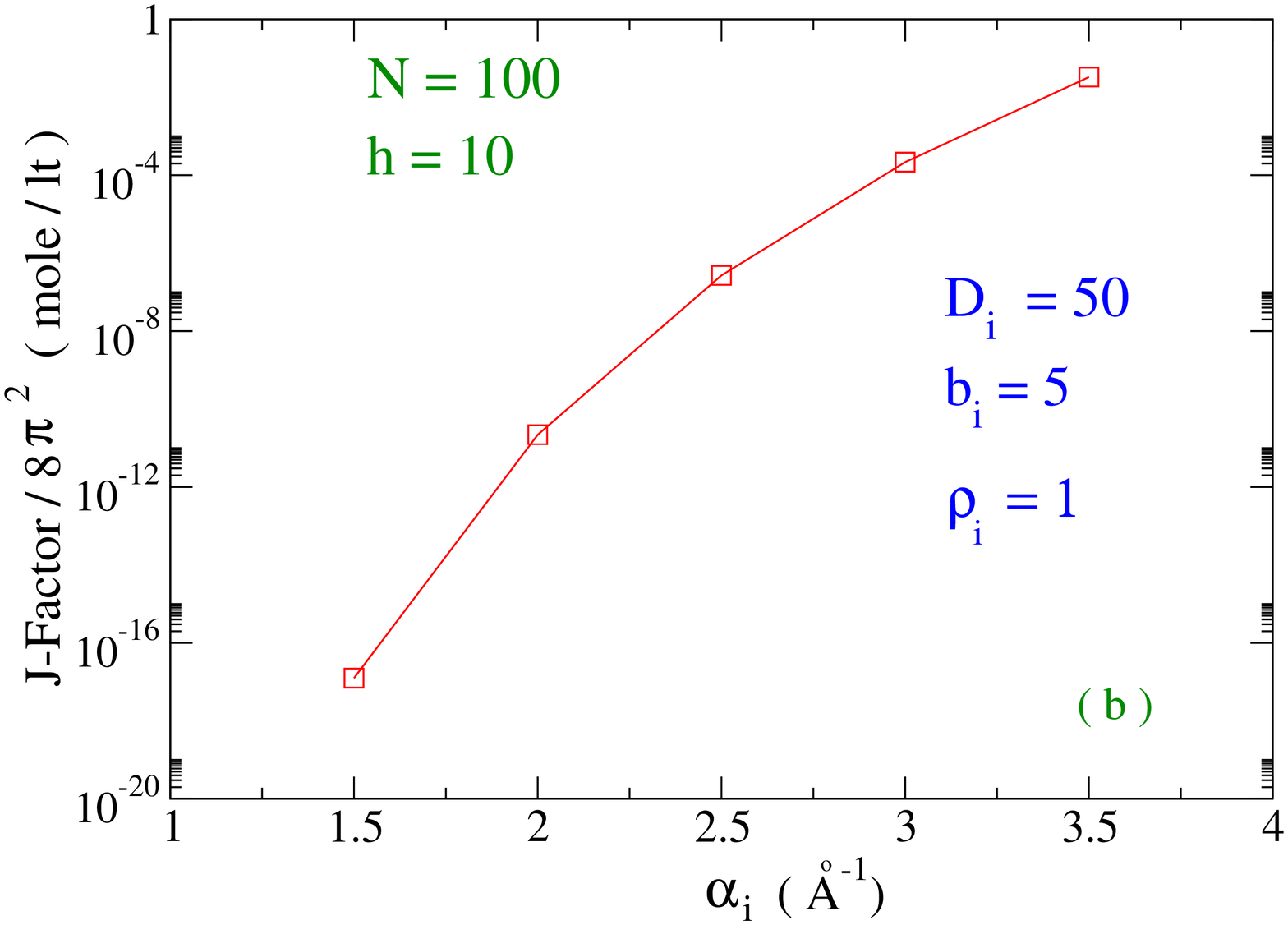}
\caption{\label{fig:4}(Color online)  
Cyclization probability, computed via Eqs.~(\ref{eq:03}),~(\ref{eq:05}), for a homogeneous molecule with $100$ base pairs and $10$ base pairs per helix turn. (a) $J$- factor (over $8 \pi^2$) versus nonlinear stacking parameter $\rho_i$.  (b)  $J$- factor (over $8 \pi^2$) versus the parameter $\alpha_i$ which sets the range of the stacking potential. $D_i$ are in units $meV$; $b_i$ are in units $\AA^{-1}$. }
\end{figure}

In Figs.~\ref{fig:4}, the  $J$- factor is plotted for a homogeneous chain with $N=\,100$\, taking the hydrogen bond parameters of Figs.~\ref{fig:3} for $GC$-\textit{bps} and, instead, varying the nonlinear stacking parameters. 

{Increasing $\rho _i$, see Fig.~\ref{fig:4}(a), yields a strong intra-strand coupling which stabilizes the helix \cite{cooper} in the open ends conformation. 
A similar trend is obtained by reducing $\alpha _i$: this enhances the threshold above which the fluctuations can move one base pair out of the stack. Small $\alpha _i$'s tend to increase the stiffness of the molecule axis and therefore to decrease the $J$- factor as it appears in Fig.~\ref{fig:4}(b).
In this regard, the $\alpha _i^{-1}$'s are a measure of the molecule persistence length although, being $V_{2}$ non linear,  the persistence length is not directly related to the rigidity parameters as in elastic models  \cite{crothers1}.

There is a high sensitivity of the $J$- factor both  on $\rho _i$ and on  $\alpha _i$.  Then one may consider to fit the model predictions to the available cyclization data in order to determine consistently the nonlinear stacking parameters. As the focus is here on very short sequences, we consider the cyclization of single DNA molecules yielding a $J$- factor  $\sim  10^{-9} \, mol / liter$ for $N \sim 100$  as measured by FRET. Note that, for this length, independent experiments report close values \cite{vafa,kim13} as shown in Fig.~\ref{fig:5}(b). 
Although sequence specificities, salt concentration and presence of defects may affect the precise cyclization estimate \cite{marko05,cherstvy,menon}, we take that order of magnitude to set a pair of values, e.g. $\alpha _i=\,2.5  \,\AA^{-1}$ and $\rho _i=\,1.3$ with the caveat that such choice is not unique. The latter values are used to compute the $J$- factor as a function of the molecule length as shown in Fig.~\ref{fig:5}(a). Thus, for all five $N$'s, the model potential parameters are kept constant. Furthermore, the boundary condition on the twist angle is always fulfilled so that the peculiar oscillations of the $J$- factor \cite{popov}, due to the twist rigidity of the double helix, do not occur here. Two close $\rho _i$'s are considered to remark the strong dependence of the cyclization probability on the molecule stiffness. The $J$- factor drops by decreasing $N$, markedly below $N=\,100$, in accordance with the qualitative general expectations.  
However such drop is not abrupt as predicted by the traditional worm-like-chain model as, for instance, the $J$- factor remains $\sim 10^{-11}$ at $N=\,80$.  Although the plots in Fig.~\ref{fig:5}(a) refer to homogeneous chains, the displayed trend (sequence length dependence) and the body of our results would not be altered by heterogeneity effects. Not even different choices for the pair ($\rho _i$, $\alpha _i$) would change such trend. 

\begin{figure}
\includegraphics[height=8.0cm,width=8.0cm,angle=-90]{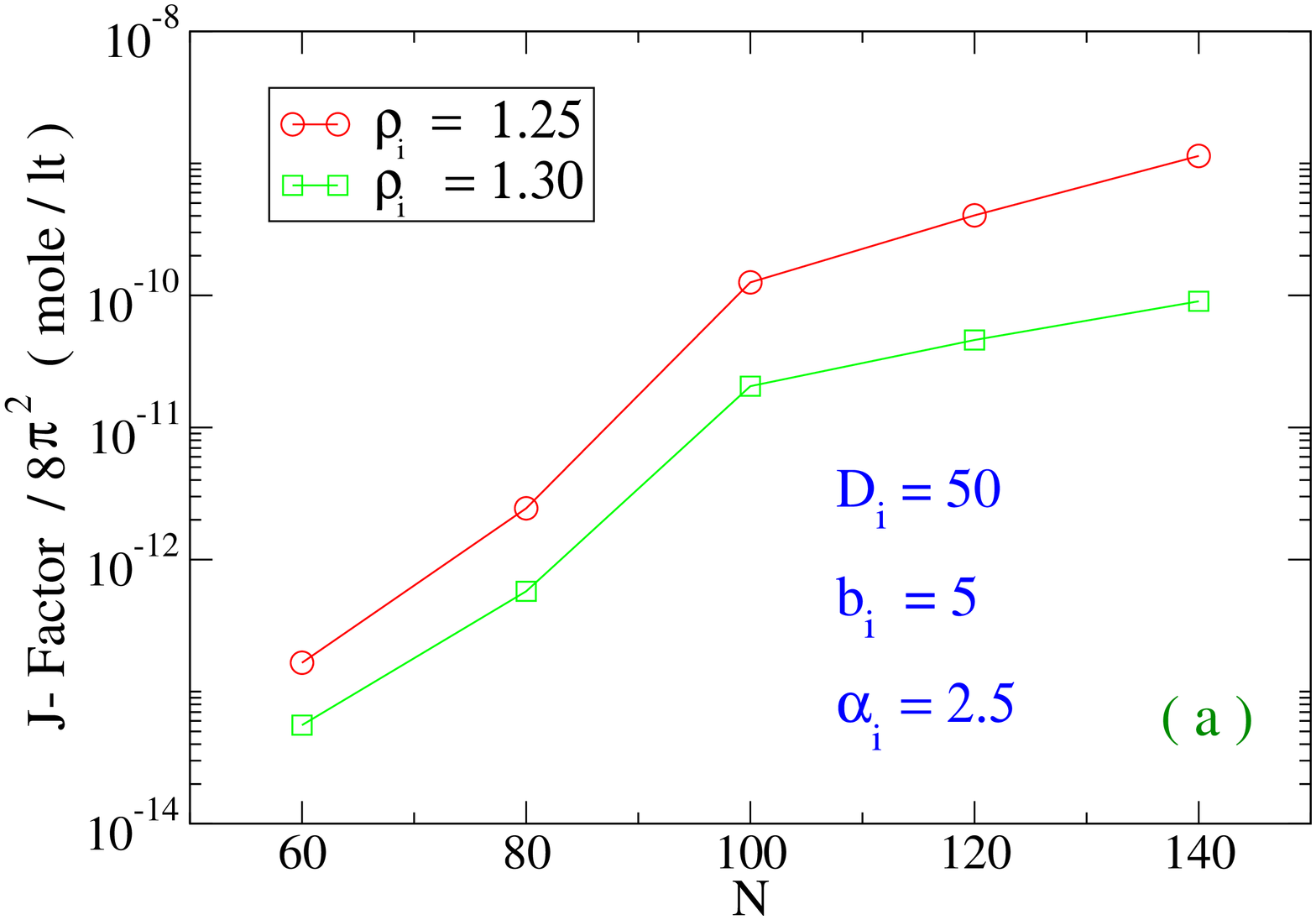}
\includegraphics[height=8.0cm,width=8.0cm,angle=-90]{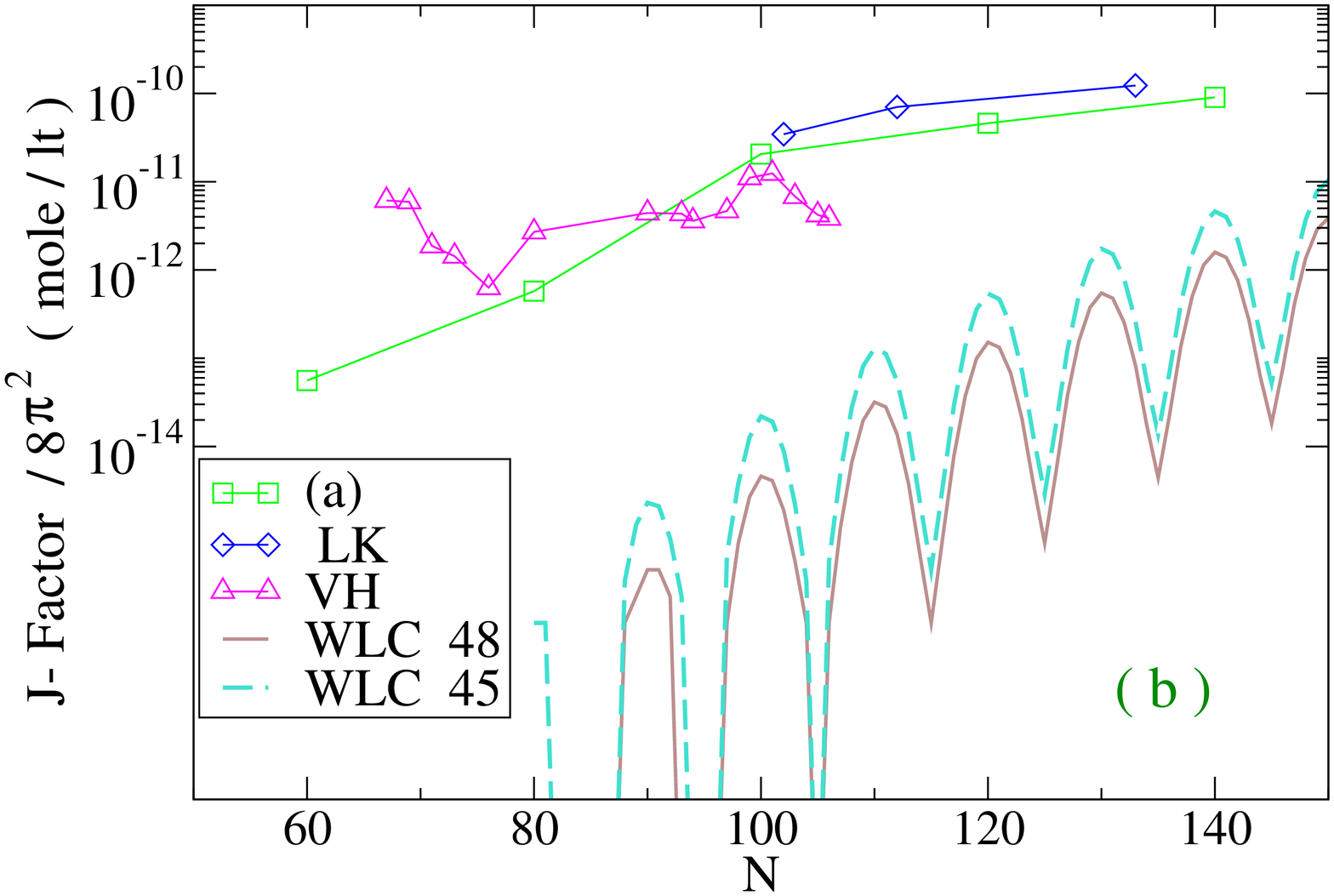}
\caption{\label{fig:5}(Color online)  
(a) $J$- factor (over $8 \pi^2$) calculated via Eq.~(\ref{eq:05}) for a set of five sequence lengths ($N$).   The potential parameters refer to homogeneous sequences and are the same for all $N$'s. Two values of the nonlinear stacking stiffness $\rho _i$ are taken. (b) The green plot in (a) is compared to the experimental results of ref.\cite{vafa} (VH) and ref.\cite{kim13} (LK). The $J$-factor of the twisted worm-like-chain model (WLC) \cite{shimada}  is computed for two persistence lengths, $45 \,nm$ and $48 \,nm$.   }
\end{figure}

The results are compared in Fig.~\ref{fig:5}(b) to FRET assays at very short length scales \cite{vafa,kim13}. There are some relevant differences between the two sets of data: \textit{i)} the molecule lenghts given in \cite{vafa} include the $10$ \textit{bps} sticky ends, whereas the lengths in \cite{kim13} do not. See also ref. \cite{kim14}.  \textit{ii)} Ref.\cite{vafa} reports an apparent $J$-factor with looping rate $R$ which is in fact the sum of the looping and unlooping terms (whose relative weight depends on salt concentration and likely on sequence length). After subtracting the unlooping contribution (Fig.3(B) in \cite{vafa}), the real $J$-factor should be about a factor three smaller for the $N=\,99$ molecule. \textit{iii)} In ref. \cite{kim13}, the annealing rates are measured with single stranded rather than double stranded molecules: although both techniques appear legitimate, this may have yielded $J$-factors values about three fold higher \cite{pc} than in ref. \cite{vafa}.
All these features combine to indicate the difficulty in extracting $J$- factors from experiments  and in performing quantitative comparison between models and data \cite{towles}. Nevertheless, also after reducing both sets of FRET data by about a factor three, the fact remains that there is a sizeable cyclization probability at very short molecule lengths. {Our calculation can predict this behavior for a consistent choice of the model parameters although the oscillations in the $J$-factor experimental plot are not reproduced for the reasons explained above. Both the structure of the stacking potential which allows for even large bending angles and the specific integration technique which allows for a broad ensemble of independent path fluctuations at any site, contribute to shape a model for the helix with flexible hinges at the level of the base pair. These mechanisms are responsible for the substantial molecule bendability which leads to the results shown in Fig.~\ref{fig:5}.
{Certainly a stricter comparison between experiments and model could be performed by accounting also for the sequence specificities of the fragments, not included in the present computation.
}
The results of the twisted WLC theory by Shimada and Yamakawa are also reported in Fig.~\ref{fig:5}(b) for two values of persistence length: the $J$- factor vanishes in the limit of short molecule lengths  but, interestingly, appreciable values would be recovered for sequences whose persistence length had to be sufficiently small. This hints to a possible way to bridge the gap, at least partly, between WLC theory and experiments. Accurate measurements of persistence length  may help to establish whether extended WLC models have predictive capability for very short molecules.

\section*{VII. Conclusions }

{We} have developed a quantitative analysis of the DNA cyclization probability, a sensitive measure of the bending and twisting of DNA molecules in solution under the effect of thermal fluctuations. The study is based on a mesoscopic Hamiltonian that models the essential interactions stabilizing the double helix and, mostly, it accounts for the  bending angles between adjacent nucleotides along the backbone of short molecules.
The computation employs a path integral method that treats the relative base pair displacements as temperature dependent trajectories and generates a large ensemble of molecule configurations in the path phase space. Boundary conditions on the path trajectories of the molecule chain ends are implemented in the numerical program to obtain the partition function for the subset of closed molecules.  The fundamental lengths of the helix,  diameter and rise distance, together with the number of base pairs per helix turn are kept fixed in the calculation whereas the model parameters are tuned to obtain free energies per base pair comparable with the experiments. In particular, after setting the values for the hydrogen bond parameters which regulate the inter-strand interactions, {we} have computed the cyclization probability, i.e., the $J$-factor, as a function of the nonlinear intra-strand potential and found a high sensitivity on the stacking parameters. Fitting the latter to the $J$-factor's order of magnitude recently estimated for a chain with $N \sim 100$ base pairs,  the computation has been extended to homogeneous molecules in the range $N \sim 60 - 140$ and found that the $J$-factor drops for $N < 100$ due to the increasingly high bending cost which are expected to hinder the cyclization in very short molecular chains. Nonetheless, even for such short lengths, the calculated $J$-factors remain sizeable {in agreement with recent single-
molecule fluorescence resonance energy transfer assays and at variance with twisted worm-like-chain models.} These findings suggest that a significant flexibility persists at scales of a few base pairs due to both large bending angles and anharmonic elasticity of the molecule axis.
While these results rely on a specific, albeit widely used, stacking potential and alternative choices could be tested, the current analysis shows that models at the mesoscopic scale can yield accurate insight into the stability parameters of the double helix provided that base pair fluctuations for the ensemble of molecule configurations are fully incorporated in the computational method.

\section*{Acknowledgements }

I wish to thank Drs. H.D. Kim,  T.T. Lee and R. Vafabakhsh for sharing details of their works.


\begin{thebibliography}{widest-label}
\bibitem{note1}
If $N / h$ is not an integer, an extra twist is required to join the strand ends. This torsional stress may become relevant in short chains thus reducing the cyclization probability. This determines the observed oscillations in the $J$- factor of small molecules.
\end{thebibliography}
\end{document}